\documentclass[twocolumn,floatfix,,superscriptaddress,showpacs]{revtex4}
\usepackage{graphicx}
\begin{document}
\title{Antibunched photons emitted by a quantum point contact out of
equilibrium}
\author{C. W. J. Beenakker}
\affiliation{Instituut-Lorentz, Universiteit Leiden, P.O. Box 9506, 2300 RA
Leiden, The Netherlands}
\author{H. Schomerus}
\affiliation{Max-Planck-Institut f\"{u}r Physik komplexer Systeme,
N\"{o}thnitzer Str.\ 38, 01187 Dresden, Germany}
\begin{abstract}
Motivated by the experimental search for ``GHz nonclassical light'', we
identify the conditions under which current fluctuations in a narrow
constriction generate sub-Poissonian radiation. Antibunched electrons
generically produce bunched photons, because the same photon mode can be
populated by electrons decaying independently from a range of initial energies.
Photon antibunching becomes possible at frequencies close to the applied
voltage $V\times e/\hbar$, when the initial energy range of a decaying electron
is restricted. The condition for photon antibunching in a narrow frequency
interval below $eV/\hbar$ reads
$[\sum_{n}T_{n}(1-T_{n})]^{2}<2\sum_{n}[T_{n}(1-T_{n})]^{2}$, with $T_{n}$ an
eigenvalue of the transmission matrix. This condition is satisfied in a quantum
point contact, where only a single $T_{n}$ differs from 0 or 1. The photon
statistics is then a superposition of binomial distributions.
\end{abstract}
\maketitle

In a recent experiment \cite{Gab04}, Gabelli et al.\ have measured the
deviation from Poisson statistics of photons emitted by a resistor in
equilibrium at mK temperatures. By cross-correlating the power fluctuations
they detected photon bunching, meaning that the variance ${\rm Var}\,n=\langle
n^{2}\rangle-\langle n\rangle^{2}$ in the number of detected photons exceeds
the mean photon count $\langle n\rangle$. Their experiment is a variation on
the quantum optics experiment of Hanbury Brown and Twiss \cite{Han56}, but now
at GHz frequencies.

In the discussion of the implications of their novel experimental technique,
Gabelli et al.\ noticed that a general theory \cite{Bee01} for the radiation
produced by a conductor out of equilibrium implies that the deviation from
Poisson statistics can go either way: Super-Poissonian fluctuations (${\rm
Var}\,n>\langle n\rangle$, signaling bunching) are the rule in conductors with
a large number of scattering channels, while sub-Poissonian fluctuations (${\rm
Var}\,n<\langle n\rangle$, signaling antibunching) become possible in
few-channel conductors. They concluded that a quantum point contact could
therefore produce GHz nonclassical light \cite{Man95}.

It is the purpose of this work to identify the conditions under which
electronic shot noise in a quantum point contact can generate antibunched
photons. The physical picture that emerges differs in one essential aspect from
electron-hole recombination in a quantum dot or quantum well, which is a
familiar source of sub-Poissonian radiation \cite{Kim99,Mic00,Yua02}. In those
systems the radiation is produced by transitions between a few discrete levels.
In a quantum point contact the transitions cover a continuous range of energies
in the Fermi sea. As we will see, this continuous spectrum generically prevents
antibunching, except at frequencies close to the applied voltage.

\begin{figure}
\includegraphics[width=8cm]{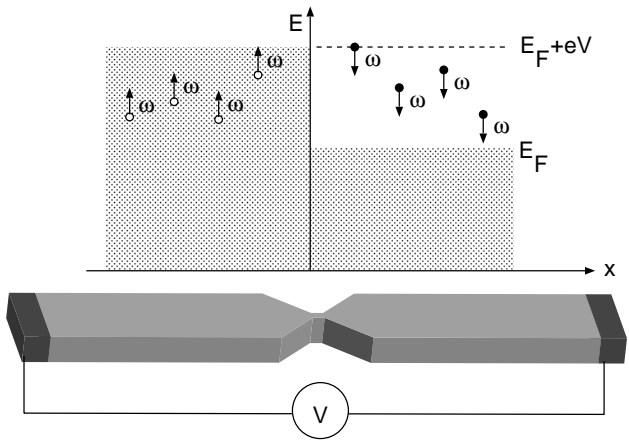}
\caption{
Schematic diagram of a constriction in a conductor (bottom) and the energy
range of electronic states (top), showing excitations of electrons (black dots)
and holes (white dots) in the Fermi sea. A voltage $V$ drops over the
constriction. Electrons (holes) in an energy range $eV-\hbar\omega$ can
populate a photon mode of frequency $\omega$, by decaying to an empty (filled)
state closer to the Fermi level.
\label{bunching}
}
\end{figure}

Before presenting a quantitative analysis, we first discuss the mechanism in
physical terms. As depicted in Fig.\ \ref{bunching}, electrons are injected
through a constriction in an energy range $eV$ above the Fermi energy $E_{F}$,
leaving behind holes at the same energy. The statistics of the charge $Q$
transferred in a time $\tau\gg\hbar/eV$ is binomial \cite{Lev93}, with ${\rm
Var}\,Q/e<\langle Q/e\rangle$. This electron antibunching is a result of the
Pauli principle. Each scattering channel $n=1,2,\ldots N$ in the constriction
and each energy interval $\delta E=\hbar/\tau$ contributes independently to the
charge statistics. The photons excited by the electrons would inherit the
antibunching if there would be a one-to-one correspondence between the transfer
of an electron and the population of a photon mode. Generically, this is not
what happens: A photon of frequency $\omega$ can be excited by each scattering
channel and by a range $eV-\hbar\omega$ of initial energies. The resulting
statistics of photocounts is negative binomial \cite{Bee01}, with ${\rm
Var}\,n>\langle n\rangle$. This is the same photon bunching as in black-body
radiation \cite{note0}.

In order to convert antibunched electrons into antibunched photons, it is
sufficient to ensure a one-to-one correspondence between electron modes and
photon modes. This can be realized by concentrating the current fluctuations in
a single scattering channel and by restricting the energy range
$eV-\hbar\omega$. Indeed, in a single-channel conductor and in a narrow
frequency range $\omega\alt eV/\hbar$ we obtain sub-Poissonian photon
statistics regardless of the value of the transmission probability. In the more
general multi-channel case, photon antibunching is found if
$[\sum_{n}T_{n}(1-T_{n})]^{2}<2\sum_{n}[T_{n}(1-T_{n})]^{2}$ (with $T_{n}$ an
eigenvalue of the transmission matrix product $tt^{\dagger}$).

Starting point of our quantitative analysis is the general relationship of
Ref.\ \cite{Bee01} between the photocount distribution $P(n)$ and the
expectation value of an ordered exponential of the electrical current operator:
\begin{eqnarray}
&&P(n)=\frac{1}{n!}\lim_{\xi\rightarrow -1}\frac{d^{n}}{d\xi^{n}}F(\xi),
\label{PFrelation}\\
&&F(\xi)=\left\langle{\cal O}\exp\left[\xi\int_{0}^{\infty}
d\omega\,\gamma(\omega)
I^{\dagger}(\omega)I(\omega)\right]\right\rangle.\label{Fresult}
\end{eqnarray}

We summarize the notation. The function
$F(\xi)=\sum_{k=0}^{\infty}(\xi^{k}/k!)\langle n^{k}\rangle_{\rm f}$ is the
generating function of the factorial moments $\langle n^{k}\rangle_{\rm
f}\equiv\langle n(n-1)(n-2)\cdots(n-k+1)\rangle$. The current operator
$I=I_{\rm out}-I_{\rm in}$ is the difference of the outgoing current $I_{\rm
out}$ (away from the constriction) and the incoming current $I_{\rm in}$
(toward the constriction). The symbol $\cal O$ indicates ordering of the
current operators from left to right in the order $I_{\rm in}^{\dagger},I_{\rm
out}^{\dagger},I_{\rm out}^{\vphantom\dagger},I_{\rm in}^{\vphantom\dagger}$.
The real frequency-dependent response function $\gamma(\omega)$ is proportional
to the coupling strength of conductor and photodetector and proportional to the
detector efficiency. Positive (negative) $\omega$ corresponds to absorption
(emission) of a photon by the detector. We consider photodetection by
absorption, hence $\gamma(\omega)\equiv 0$ for $\omega\leq 0$. Integrals over
frequency should be interpreted as sums over discrete modes $\omega_{p}=p\times
2\pi/\tau$, $p=1,2,3,\ldots$. The detection time $\tau$ is sent to infinity at
the end of the calculation. We denote $\gamma_{p}=\gamma(\omega_{p})\times
2\pi/\tau$, so that $\int
d\omega\,\gamma(\omega)\rightarrow\sum_{p}\gamma_{p}$. For ease of notation we
set $\hbar=1$, $e=1$.

The exponent in Eq.\ (\ref{Fresult}) is quadratic in the current operators,
which complicates the calculation of the expectation value. We remove this
complication by introducing a Gaussian field $z(\omega)$ and performing a
Hubbard-Stratonovich transformation,
\begin{eqnarray}
F(\xi)&=&\left\langle{\cal O}\exp\left[\sqrt{\xi}\int_{0}^{\infty}
d\omega\,\gamma(\omega) [z(\omega)I^{\dagger}(\omega)\right.\right.\nonumber\\
&&\mbox{}+\left.\left.\vphantom{\int_{0}^{\infty}}z^{\ast}(\omega)I(\omega)]
\right]\right\rangle.\label{FHS}
\end{eqnarray}
The brackets $\langle\cdots\rangle$ now indicate both a quantum mechanical
expectation value of the current operators and a classical average over
independent complex Gaussian variables $z_{p}=z(\omega_{p})$ with zero mean and
variance $\langle |z_{p}|^{2}\rangle=1/\gamma_{p}$.

We assume zero temperature, so that the incoming current is noiseless. We may
then replace $I$ by $I_{\rm out}$ and restrict ourselves to energies
$\varepsilon$ in the range $(0,V)$ above $E_{F}$. Let
$b_{n}^{\dagger}(\varepsilon)$ be the operator that creates an outgoing
electron in scattering channel $n$ at energy $\varepsilon$. The outgoing
current is given in terms of the electron operators by
\begin{equation}
I_{\rm out}(\omega)=\int_{0}^{V}
d\varepsilon\,\sum_{n}b^{\dagger}_{n}(\varepsilon)
b^{\vphantom{\dagger}}_{n}(\varepsilon+\omega).\label{Iout}
\end{equation}
Energy $\varepsilon_{p}=p\times 2\pi/\tau$ is discretized in the same way as
frequency. The energy and channel indices $p,n$ are collected in a vector $b$
with elements $b_{pn}=(2\pi/\tau)^{1/2}b_{n}(\epsilon_{p})$. Substitution of
Eq.\ (\ref{Iout}) into Eq.\ (\ref{FHS}) gives
\begin{equation}
F(\xi)=\left\langle e^{b^{\dagger}Zb}e^{b^{\dagger}Z^{\dagger}b}\right\rangle.
\label{FZ}
\end{equation}
The exponents contain the product of the vectors $b,b^{\dagger}$ and a matrix
$Z$ with elements $Z_{pn,p'n'}=\xi^{1/2}\delta_{nn'}z_{p-p'}\gamma_{p-p'}$.
Notice that $Z$ is diagonal in the channel indices $n,n'$ and lower-triangular
in the energy indices $p,p'$.

Because of the ordering ${\cal O}$ of the current operators, the single
exponential of Eq.\ (\ref{FHS}) factorizes into the two noncommuting
exponentials of Eq.\ (\ref{FZ}). In order to evaluate the expectation value
efficiently, we would like to bring this back to a single exponential --- but
now with normal ordering ${\cal N}$ of the fermion creation and annihilation
operators. (Normal ordering means $b^{\dagger}$ to the left of $b$, with a
minus sign for each permutation.) This is accomplished by means of the operator
identity \cite{note1}
\begin{equation}
\prod_{i}e^{b^{\dagger}A_{i}b}={\cal
N}\exp\left[b^{\dagger}\left(\prod_{i}e^{A_{i}}-1\right)b\right],
\label{normalordering}
\end{equation}
valid for any set of matrices $A_{i}$. The quantum mechanical expectation value
of a normally ordered exponential is a determinant \cite{Cah99},
\begin{equation}
\langle {\cal N}e^{b^{\dagger}Ab}\rangle={\rm Det}\,(1+AB),\;\; B_{ij}=\langle
b_{j}^{\dagger}b_{i}^{\vphantom{\dagger}}\rangle.\label{determinant}
\end{equation}
In our case $A=e^{Z}e^{Z^{\dagger}}-1$ and $B=tt^{\dagger}$, with $t$ the
$N\times N$ transmission matrix of the constriction.

In the experimentally relevant case \cite{Agu00,Gab04} the response function
$\gamma(\omega)$ is sharply peaked at a frequency $\Omega\alt V$, with a width
$\Delta\ll\Omega$. We assume that the energy dependence of the transmission
matrix may be disregarded on the scale of $\Delta$, so that we may choose an
$\varepsilon$-independent basis in which $tt^{\dagger}$ is diagonal. The
diagonal elements are the transmission eigenvalues $T_{1},T_{2},\ldots T_{N}\in
(0,1)$. Combining Eqs.\ (\ref{FZ}--\ref{determinant}) we arrive at
\begin{eqnarray}
F(\xi)&=&\left\langle\prod_{n=1}^{N}{\rm
Det}\,\left[1+T_{n}\bigl(e^{Z}e^{Z^{\dagger}}-1\bigr)\right]\right\rangle
\nonumber\\
&=&\left\langle\prod_{n=1}^{N}{\rm
Det}\,\left[(1-T_{n})e^{-Z^{\dagger}}+T_{n}e^{Z}\right]\right\rangle.
\label{Fxieindependent}
\end{eqnarray}
(In the second equality we used that ${\rm Det}\,e^{Z^{\dagger}}=1$, since $Z$
is a lower-triangular matrix.) The remaining average is over the Gaussian
variables $z_{p}$ contained in the matrix $Z$.

Since the interesting new physics occurs when $\Omega$ is close to $V$, we
simplify the analysis by assuming that $\gamma(\omega)\equiv 0$ for
$\omega<V/2$. For such a response function one has $Z^{2}=0$. (This amounts to
the statement that no electron with excitation energy $\varepsilon<V$ can
produce more than a single photon of frequency $\omega>V/2$.) We may therefore
replace $e^{Z}\rightarrow 1+Z$ and $e^{-Z^{\dagger}}\rightarrow 1-Z^{\dagger}$
in Eq.\ (\ref{Fxieindependent}). We then apply the matrix identity
\begin{equation}
{\rm Det}\,(1+A+B)={\rm Det}\,(1-AB),\;\;{\rm
if}\;\;A^{2}=0=B^{2},\label{Detidentity}
\end{equation}
and obtain
\begin{eqnarray}
F(\xi)&=&\prod_{p}\frac{\gamma_{p}}{\pi}\int
d^{2}z_{p}\,e^{-\gamma_{p}|z_{p}|^{2}}\nonumber\\
&&\mbox{}\times\prod_{n=1}^{N}{\rm Det}\,[1+T_{n}(1-T_{n})\xi X].\label{FxiTT}
\end{eqnarray}
We have defined $\xi X\equiv ZZ^{\dagger}$ and written out the Gaussian
average. The Hermitian matrix $X$ has elements
\begin{equation}
X_{pp'}=\sum_{q}z^{\vphantom{\ast}}_{p-q}z^{\ast}_{p'-q}\gamma_{p-q}\gamma_{p'-q}. 
\label{Xdef}
\end{equation}
The integers $p,p',q$ range from $1$ to $V\tau/2\pi$.

The Gaussian average is easy if the dimensionless shot noise power
$S=\sum_{n}T_{n}(1-T_{n})$ is $\gg 1$. We may then do the integrals of Eq.\
(\ref{FxiTT}) in saddle-point approximation, with the result \cite{note2}
\begin{equation}
\ln F(\xi)=-\frac{\tau}{2\pi}\int_{0}^{V}d\omega\,\ln\left[1-\xi
S\gamma(\omega)(V-\omega)\right].\label{Fxisaddle}
\end{equation}
The logarithm $\ln F(\xi)$ is the generating function of the factorial
cumulants $\langle\!\langle n^{k}\rangle\!\rangle_{\rm f}$ \cite{note3}. By
expanding Eq.\ (\ref{Fxisaddle}) in powers of $\xi$ we find
\begin{equation}
\langle\!\langle n^{k}\rangle\!\rangle_{\rm
f}=(k-1)!\frac{\tau}{2\pi}\int_{0}^{V}d\omega\,[S\gamma(\omega)(V-\omega)]^{k}.
\label{nfactsaddlepoint}
\end{equation}
Eqs.\ (\ref{Fxisaddle}) and (\ref{nfactsaddlepoint}) represent the multi-mode
superposition of independent negative-binomial distributions \cite{note0}. All
factorial cumulants are positive, in particular the second, so ${\rm
Var}\,n>\langle n\rangle$. This is super-Poissonian radiation.

When $S$ is not $\gg 1$, e.g.\ when only a single channel contributes to the
shot noise, the result (\ref{Fxisaddle}-\ref{nfactsaddlepoint}) remains valid
if $V-\Omega\gg\Delta$. This was the conclusion of Ref.\ \cite{Bee01}, that
narrow-band detection leads generically to a negative-binomial distribution.
However, the saddle-point approximation breaks down when the detection
frequency $\Omega$ approaches the applied voltage $V$. For $V-\Omega\alt\Delta$
one has to calculate the integrals in Eq.\ (\ref{FxiTT}) exactly.

We have evaluated the generating function (\ref{FxiTT}) for a response function
of the block form
\begin{equation}
\gamma(\omega)=\left\{\begin{array}{l}
\gamma_{0}\;\;{\rm if}\;\;V-\Delta<\omega<V,\\
0\;\;{\rm if}\;\;\omega<V-\Delta,
\end{array}\right.\label{gammadef}
\end{equation}
with $\Delta<V/2$. The frequency dependence for $\omega>V$ is irrelevant. In
the case $N=1$ of a single channel, with transmission probability $T_{1}\equiv
T$, we find \cite{note4}
\begin{eqnarray}
\ln F(\xi)&=&\frac{\tau}{2\pi}\int_{V-\Delta}^{V}d\omega\,\ln\left[1+\xi
\gamma_{0}T(1-T)(V-\omega)\right]\nonumber\\
&=&\frac{\tau\Delta}{2\pi}\frac{(1+x)\ln(1+x)-x}{x},\label{Fxiexact}
\end{eqnarray}
with $x\equiv\xi\gamma_{0} T(1-T)\Delta$. This is a superposition of binomial
distributions. The factorial cumulants are
\begin{equation}
\langle\!\langle n^{k}\rangle\!\rangle_{\rm
f}=(-1)^{k+1}\frac{(k-1)!}{k+1}\frac{\tau\Delta}{2\pi}[T(1-T)\gamma_{0}\Delta]^{k}.
\label{nfactexact}
\end{equation}
The second factorial cumulant is negative, so ${\rm Var}\,n<\langle n\rangle$.
This is sub-Poissonian radiation.

We have not found such a simple closed-form expression in the more general
multi-channel case, but it is straightforward to evaluate the low-order
factorial cumulants from Eq.\ (\ref{FxiTT}). We find
\begin{eqnarray}
&&\langle n\rangle=\frac{\tau\Delta}{2\pi}\gamma_{0}\Delta\frac{1}{2}S_{1},
\label{n1}\\
&&\langle\!\langle n^{2}\rangle\!\rangle_{\rm
f}=\frac{\tau\Delta}{2\pi}(\gamma_{0}\Delta)^{2}\frac{1}{3}(S_{1}^{2}-2S_{2}),
\label{n2}\\
&&\langle\!\langle n^{3}\rangle\!\rangle_{\rm
f}=\frac{\tau\Delta}{2\pi}(\gamma_{0}\Delta)^{3}\frac{1}{6}
\left(3S_{1}^{3}-15S_{1}S_{2}+15S_{3}\right),\label{n3}
\end{eqnarray}
with $S_{p}=\sum_{n}[T_{n}(1-T_{n})]^{p}$. Antibunching therefore requires
$S_{1}^{2}<2S_{2}$.

The condition on antibunching can be generalized to arbitrary frequency
dependence of the response function $\gamma(\omega)$ in the range
$V-\Delta<\omega<V$ of detected frequencies. For $\Delta<V/2$ we find
\begin{eqnarray}
&&{\rm Var}\,n-\langle
n\rangle=\frac{\tau}{2\pi}\int_{V-\Delta}^{V}d\omega'\,\gamma(\omega')
\int_{\omega'}^{V}d\omega\,(V-\omega)\nonumber\\
&&\;\;\;\mbox{}\times\left(2S_{1}^{2}-4S_{2}-(V-\omega)S_{1}^{2}
\frac{d}{d\omega}\right) \gamma(\omega).\label{Varngeneral}
\end{eqnarray}
We see that the antibunching condition $S_{1}^{2}<2S_{2}$ derived for the
special case of the block function (\ref{gammadef}) is more generally a
sufficient condition for antibunching to occur, provided that
$d\gamma/d\omega\geq 0$ in the detection range. It does not matter if the
response function drops off at $\omega>V$, provided that it increases
monotonically in the range $(V-\Delta,V)$. A steeply increasing response
function in this range is more favorable, but not by much. For example, the
power law $\gamma(\omega)\propto(\omega-V+\Delta)^{p}$ gives the antibunching
condition $S_{1}^{2}<2S_{2}\times[1+p/(1+p)]$, which is only weakly dependent
on the power $p$.

In conclusion, we have presented both a qualitative physical picture and a
quantitative analysis for the conversion of electron to photon antibunching. A
simple criterion, Eq.\ (\ref{n2}), is obtained for sub-Poissonian photon
statistics, in terms of the transmission eigenvalues $T_{n}$ of the conductor.
Since an $N$-channel quantum point contact has only a single $T_{N}$ different
from 0 or 1, it should generate antibunched photons in a frequency band
$(V-\Delta,V)$ --- regardless of the value of $T_{N}$. The statistics of these
photons is the superposition (\ref{Fxiexact}) of binomial distributions,
inherited from the electronic binomial distribution. There are no stringent
conditions on the band width $\Delta$, as long as it is $<V/2$ (in order to
prevent multi-photon excitations by a single electron \cite{note5}). This
should make it feasible to use the cross-correlation technique of Ref.\
\cite{Gab04} to detect the emission of nonclassical microwaves by a quantum
point contact.

We have benefitted from correspondence with D. C. Glattli. This work was
supported by the Dutch Science Foundation NWO/FOM.


\begin{thebibliography}{99}
\bibitem{Gab04} J. Gabelli, L.-H. Reydellet, G. F\`{e}ve, J.-M. Berroir, B.
Pla\c{c}ais, P. Roche, and D. C. Glattli, cond-mat/0403584.
\bibitem{Han56} R. Hanbury Brown and R. Q. Twiss, Nature {\bf 177}, 27 (1956).
\bibitem{Bee01} C. W. J Beenakker and H. Schomerus, Phys.\ Rev.\ Lett.\ {\bf
86}, 700 (2001).
\bibitem{Man95} Sub-Poissonian radiation is called ``nonclassical'', because
its photocount statistics can not be interpreted in classical terms as a
superposition of Poisson processes. See L. Mandel and E. Wolf, {\em Optical
Coherence and Quantum Optics\/} (Cambridge University, Cambridge, 1995).
\bibitem{Kim99} J. Kim, O. Benson, H. Kan, and Y. Yamamoto, Nature {\bf 397},
500 (1999); C. Santori, M. Pelton, G. Solomon, Y. Dale, and Y. Yamamoto, Phys.\
Rev.\ Lett.\ {\bf 86}, 1502 (2001).
\bibitem{Mic00} P. Michler, A. Imamo\v{g}lu, M. D. Mason, P. J. Carson, G. F.
Strouse, and S. K. Buratto, Nature {\bf 406}, 968 (2000); P. Michler, A. Kiraz,
C. Becher, W. V. Schoenfeld, P. M. Petroff, L. Zhang, E. Hu, and A.
Imamo\v{g}lu, Science {\bf 290}, 2282 (2000).
\bibitem{Yua02} Z. L. Yuan, B. E. Kardynal, R. M. Stevenson, A. J. Shields, C.
J. Lobo, K. Cooper, N. S. Beattie, D. A. Ritchie, and M. Pepper, Science {\bf
295}, 102 (2002).
\bibitem{Lev93} L. S. Levitov and G. B. Lesovik, JETP Lett.\ {\bf 58}, 230
(1993).
\bibitem{note0} The {\em negative-binomial\/} distribution $P(n)\propto
{n+\nu-1\choose n} [\nu/\langle n\rangle+1]^{-n}$ counts the number of
partitions of $n$ {\em bosons\/} among $\nu=\tau\delta\omega/2\pi$ states in a
frequency interval $\delta\omega$. The {\em binomial\/} distribution
$P(n)\propto {\nu\choose n} [\nu/\langle n\rangle-1]^{-n}$ counts the number of
partitions of $n$ {\em fermions\/} among $\nu$ states.
\bibitem{note1} Eq.\ (\ref{normalordering}) is the multi-matrix generalization
of the well known identity $\exp(b^{\dagger}A b)={\cal
N}\exp[b^{\dagger}(e^{A}-1)b]$.
\bibitem{Cah99} K. E. Cahill and R. J. Glauber, Phys.\ Rev.\ A {\bf 59}, 1538
(1999).
\bibitem{Agu00} R. Aguado and L. P. Kouwenhoven, Phys.\ Rev.\ Lett.\ {\bf 84},
1986 (2000).
\bibitem{note2} The saddle point is at $z_{p}=0$, so to integrate out the
Gaussian fluctuations around the saddle point we may linearize the determinant
in Eq.\ (\ref{FxiTT}): $\prod_{n}{\rm Det}\,[1+T_{n}(1-T_{n})\xi X]=\exp[\xi
S{\rm Tr}\,X+{\cal O}(X^{2})]$. The result is Eq.\ (\ref{Fxisaddle}).
\bibitem{note3} Factorial cumulants are constructed from factorial moments in
the usual way. The first two are:  $\langle\!\langle n\rangle\!\rangle_{\rm
f}=\langle n\rangle$, $\langle\!\langle n^{2}\rangle\!\rangle_{\rm f}=\langle
n^{2}\rangle_{\rm f}-\langle n\rangle^{2}={\rm Var}\, n-\langle n\rangle$.
\bibitem{note4} Using computer algebra, we find that $\ln\langle {\rm
Det}\,[1+\xi
T(1-T)X]\rangle=\sum_{m=1}^{M}\ln[1+m\xi\gamma_{0}T(1-T)(2\pi/\tau)]$, for each
matrix dimensionality $M$ that we could check. We are confident that this
closed form holds for all $M$, but we have not yet found an analytical proof.
Eq.\ (\ref{Fxiexact}) follows in the limit $M\equiv
\tau\Delta/2\pi\rightarrow\infty$ upon conversion of the summation into an
integration.
\bibitem{note5} Multi-photon excitations do not contribute to ${\rm Var}\,n$ if
$T_{n}\in\{0,1/2,1\}$ for all $n$ [cf.\ Ref.\ \cite{Bee01}, Eq.\ (19)]. For a
quantum point contact, one finds that antibunching persists when $\Delta>V/2$
provided that $T_{N}(1-T_{N})>1/6$.
\end{thebibliography}
\end{document}